\documentclass[a4paper,11pt]{article}
\usepackage{pos}
\usepackage{multirow,bigdelim}  
\usepackage{multicol}

\def\red{\textcolor{red}}
\def\blue{\textcolor{blue}}
\def\green{\textcolor{ForestGreen}}

\title{Probing the Earth's Core using Atmospheric Neutrinos at INO}

\author[a,b,c]{Anil Kumar}
\author*[a,c,d]{Sanjib Kumar Agarwalla}

\affiliation[a]{Institute of Physics, Sachivalaya Marg, Sainik School Post,
	Bhubaneswar 751005, India}
\affiliation[b]{Applied Nuclear Physics Division, Saha Institute of
	Nuclear Physics, Block AF, Sector 1, Bidhannagar, Kolkata 700064, India}
\affiliation[c]{Homi Bhabha National Institute, Anushakti Nagar,
	Mumbai 400094, India}
\affiliation[d]{International Centre for Theoretical Physics, 
	Strada Costiera 11, 34151 Trieste, Italy}

\emailAdd{anil.k@iopb.res.in (ORCID: 0000-0002-8367-8401)}
\emailAdd{sanjib@iopb.res.in (ORCID: 0000-0002-9714-8866)}

\abstract{The proposed 50 kt Iron Calorimeter (ICAL) detector at the India-based Neutrino Observatory (INO) aims to detect atmospheric muon neutrinos and antineutrinos separately in the multi-GeV range of energies and over a wide range of path lengths. While passing through the Earth, the upward-going neutrinos experience a density-dependent matter effect, which can be utilized to extract information about the internal structure of Earth. Since the Earth's matter effect modifies the neutrino oscillation patterns differently for neutrinos and antineutrinos, the capability of ICAL to distinguish $\mu^-$ and $\mu^+$ events plays an important role in observing this matter effect. Taking advantage of good angular resolution, ICAL would be able to observe about 331 $\mu^-$ and 146 $\mu^+$ events corresponding to the core-passing neutrinos and antineutrinos, respectively, in 10 years. We demonstrate for the first time that ICAL would be able to validate the presence of Earth's core by ruling out a two-layered profile consisting of only mantle and crust in fit with respect to the PREM profile in data with a median $\Delta \chi^2$ of 7.45 for normal mass ordering (NO) and 4.83 for inverted mass ordering (IO) using 500 kt$\cdot$yr exposure. If we do not use the charge identification capability of ICAL, these sensitivities deteriorate to a $\Delta\chi^2$ of 3.76 for NO and 1.59 for IO.}

\FullConference{%
  *** The European Physical Society Conference on High Energy Physics (EPS-HEP2021), ***\\
  *** 26-30 July 2021 ***\\
  *** Online conference, jointly organized by Universität Hamburg and the research center DESY ***
}

\begin{document}
\maketitle

\section{Introduction and Motivation}

Neutrinos are capable of passing through the whole Earth because they only take part in weak interactions. While traveling through the Earth, the atmospheric neutrinos interact with the ambient electrons via the charged-current (CC) process, which is coherent, forward, and elastic in nature. These CC interactions modify the neutrino and antineutrino oscillation patterns in different fashions. This Earth's matter effect~\cite{Wolfenstein:1977ue} depends on the density distribution of electrons inside Earth and can be utilized to reveal the internal structure of Earth.

From seismic studies~\cite{Dziewonski:1981xy}, we know that the Earth can mainly be divided into three layers which are core, mantle, and crust, where the core is the innermost layer. While traveling through the mantle, neutrinos experience the well-known Mikheyev-Smirnov-Wolfenstein (MSW) resonance~\cite{Wolfenstein:1977ue} around 6 to 10 GeV of energies. On the other hand, when neutrinos penetrate the core, the so-called neutrino oscillation length resonance (NOLR)~\cite{Petcov:1998su}, or parametric resonance~\cite{Akhmedov:1998ui} comes into the picture around 3 to 6 GeV of energies. 

The upcoming 50 kt Iron Calorimeter (ICAL) detector at the India-based Neutrino Observatory (INO)~\cite{Kumar:2017sdq} would detect atmospheric neutrinos and antineutrinos separately in the multi-GeV range of energies over a wide range of baselines. The magnetic field of around 1.5 Tesla enables ICAL to distinguish $\mu^-$ and $\mu^+$ events, which in turn, helps us to differentiate between the matter effects felt by neutrinos and antineutrinos. In this work, we estimate ICAL sensitivities for distinguishing various profiles of Earth using atmospheric neutrinos and antineutrinos. We validate the presence of Earth's core by ruling out a simple two-layered coreless profile of mantle-crust with respect to the PREM profile of Earth. 

\section{Studying Various Profiles of Earth through Neutrino Oscillograms}

\begin{figure}[htb!]
	\centering
	\includegraphics[width=0.33\textwidth]{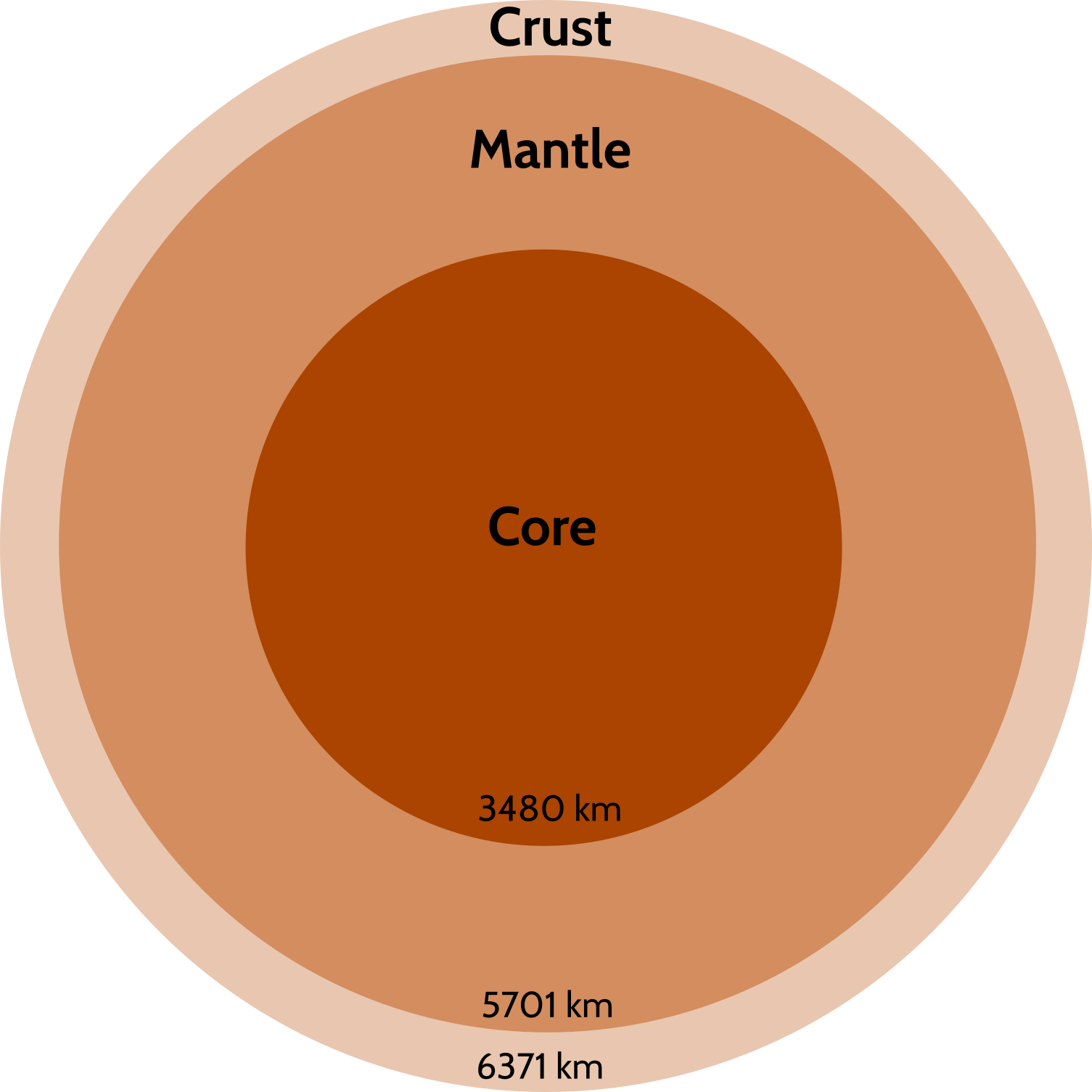} \hspace{1cm}
	\includegraphics[width=0.4\textwidth]{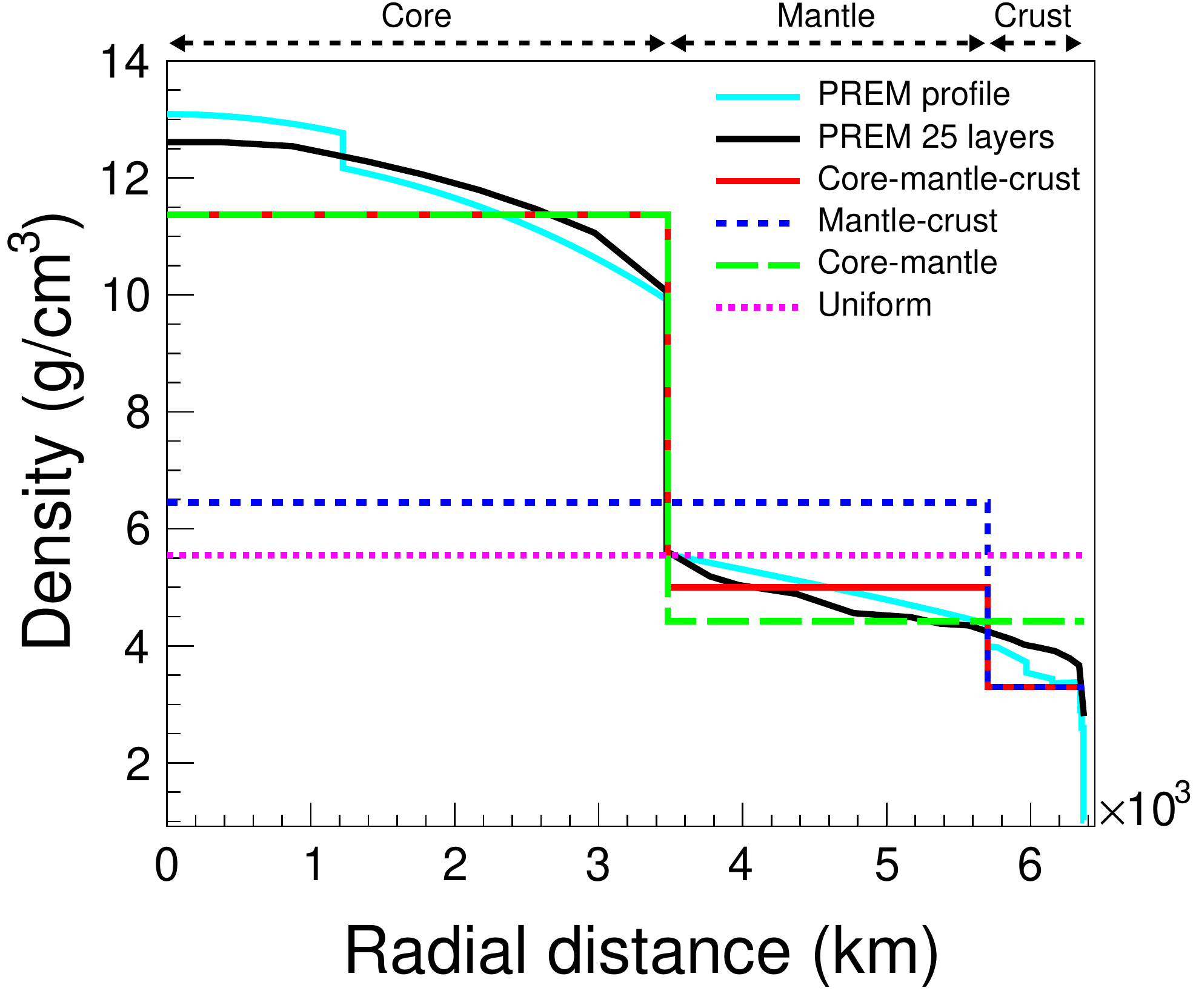}
	\caption{Left: Profile of Earth with three layers. Right: Density distributions of various profiles of Earth as a function of the radial distances of layers from the center of the Earth. Note that the total mass and radius of Earth remain invariant in all these profiles. These figures are taken from Ref.~\cite{Kumar:2021faw}.}
	\label{fig:Earth_profile}
\end{figure}

In the left panel of Fig.~\ref{fig:Earth_profile}, we show a three-layered profile of Earth with core, mantle, and crust. The right panel of Fig.~\ref{fig:Earth_profile} depicts the density distributions of various profiles of Earth as a function of the radial distances of layers from the center of Earth. The cyan curve corresponds to the PREM profile with 81 layers~\cite{Dziewonski:1981xy}. For computational ease, throughout this work, we use the PREM profile with 25 layers (black curve) which preserves all the important features of Earth. The solid red, dotted blue, dashed green, and dotted pink curves represent the core-mantle-crust, mantle-crust, core-mantle, and uniform profiles of Earth, respectively. In the coreless profile of mantle-crust, the core and mantle are merged together, having a  uniform density throughout the resulting mantle. While considering all these profiles, we assume that the mass and radius of Earth remain invariant\footnote{Note that one can also consider the moment of inertia of Earth as an additional invariant quantity.}.

\begin{figure}[t]
	\centering
	\includegraphics[width=0.4\linewidth]{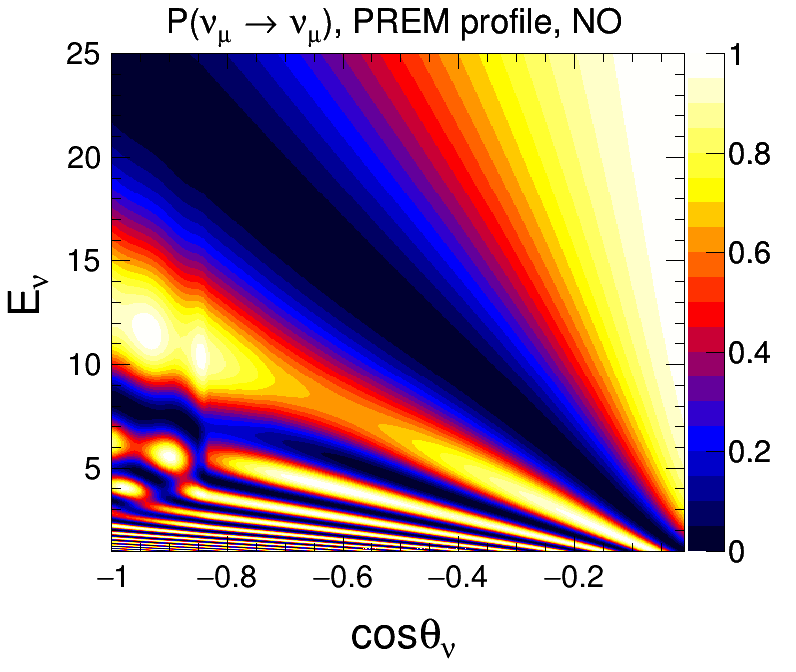} \hspace{0.5cm}
	\includegraphics[width=0.4\linewidth]{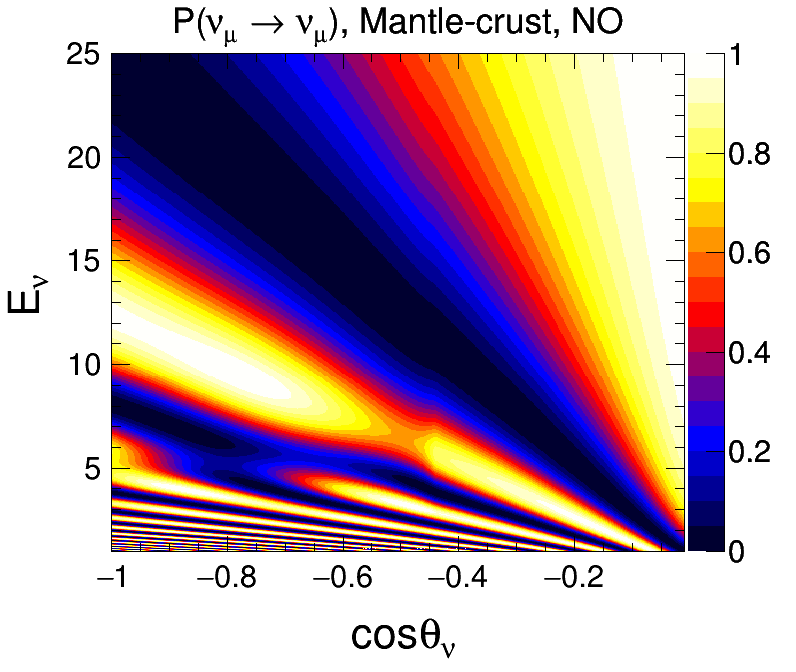}
	\caption{Oscillograms for $\nu_\mu$ survival channel considering the PREM profile (left panel) and two-layered mantle-crust profile (right panel) of the Earth. See text for details. These figures are taken from Ref.~\cite{Kumar:2021faw}.}
	\label{fig:puu-nu-mat25}
\end{figure}

\begin{table}[htb!]
	\centering
	\begin{tabular}{|c|c|c|c|c|c|c|}
		\hline
		$\sin^2 2\theta_{12}$ & $\sin^2\theta_{23}$ & $\sin^2 2\theta_{13}$ & $\Delta m^2_
		\text{eff}$ (eV$^2$) & $\Delta m^2_{21}$ (eV$^2$) & $\delta_{\rm CP}$ & Mass Ordering\\
		\hline
		0.855 & 0.5 & 0.0875 & $2.49\times 10^{-3}$ & $7.4\times10^{-5}$ & 0 & Normal (NO)\\
		\hline 
	\end{tabular}
	\caption{The values of benchmark oscillation parameters used in this analysis. These values are in good agreement with the present neutrino global fits~\cite{NuFIT}.  Normal mass ordering corresponds to $m_1 < m_2 < m_3$.}
	\label{tab:osc-param-value}
\end{table}

About 98\% events at ICAL are contributed by survival $(\nu_\mu \rightarrow \nu_\mu)$ channel and remaining by appearance $(\nu_e \rightarrow \nu_\mu)$ channel. Hence, in Fig.~\ref{fig:puu-nu-mat25}, we only show oscillograms for $\nu_\mu$ survival channel as a function of neutrino energies and directions for the PREM profile (see left panel) and mantle-crust profile (see right panel) using the oscillation parameters mentioned in Table~\ref{tab:osc-param-value}. $\Delta m^2_\text{eff}$ is the effective atmospheric mass squared difference\footnote{$\Delta m^2_\text{eff}$ is related to $\Delta m^2_{31}$ by $\Delta m^2_\text{eff} = \Delta m^2_{31} - \Delta m^2_{21} (\cos^2\theta_{12} - \cos \delta_\text{CP} \sin\theta_{13}\sin2\theta_{12}\tan\theta_{23}).$}.
The dark blue diagonal bands in both the panels of Fig.~\ref{fig:puu-nu-mat25} correspond to the minima of $\nu_\mu \rightarrow \nu_\mu$ survival channel and are known as the oscillation valleys~\cite{Kumar:2020wgz,Kumar:2021lrn}. In the left panel of Fig.~\ref{fig:puu-nu-mat25}, the MSW resonance can be seen as red patch around $-0.8 < \cos\theta_\nu < -0.5$ and  $6 {\rm~GeV} < E_\nu < 10 {\rm~GeV}$, whereas yellow patches around $-1.0 < \cos\theta_\nu < -0.8$ and $3 {\rm~GeV} < E_\nu < 6 {\rm~GeV}$ are caused by NOLR/parametric resonance. As far as the mantle-crust profile is concerned, we can see in the right panel of Fig.~\ref{fig:puu-nu-mat25} that the MSW resonance region gets modified and the parametric resonance zone is absent which indicates that the core is not there anymore.

\section{Reconstructed Muon Event Distributions at ICAL}

The magnetized 50 kt ICAL detector consists of stacks of iron layers (act as targets) with Resistive Plate Chambers (RPCs) sandwiched between them as active elements. The charged-current quasi-elastic interactions of atmospheric muon neutrinos inside the iron layers of ICAL produce only muons in the final state, which create hits in the RPCs, resulting in long tracks inside the detector. The direction of bending of the muon track in the presence of the magnetic field gives rise to the charge identification (CID) capability of ICAL. On the other hand, the charged-current resonance scatterings and deep inelastic interactions create hadron showers along with muon tracks. 

We simulate neutrino interactions using the NUANCE event generator considering ICAL geometry with neutrino flux at the INO site. We incorporate the full three-flavor neutrino oscillation probabilities using a reweighting algorithm. The detector responses are applied using the migration matrices provided by the ICAL collaboration~\cite{Kumar:2017sdq}. To suppress the statistical fluctuations, first, we generate the events considering 1000-yr exposure that is later scaled to 10-yr Monte Carlo (MC) data corresponding to 500 kt$\cdot$yr exposure.

\begin{figure}
	\centering
	\includegraphics[width=0.4\linewidth]{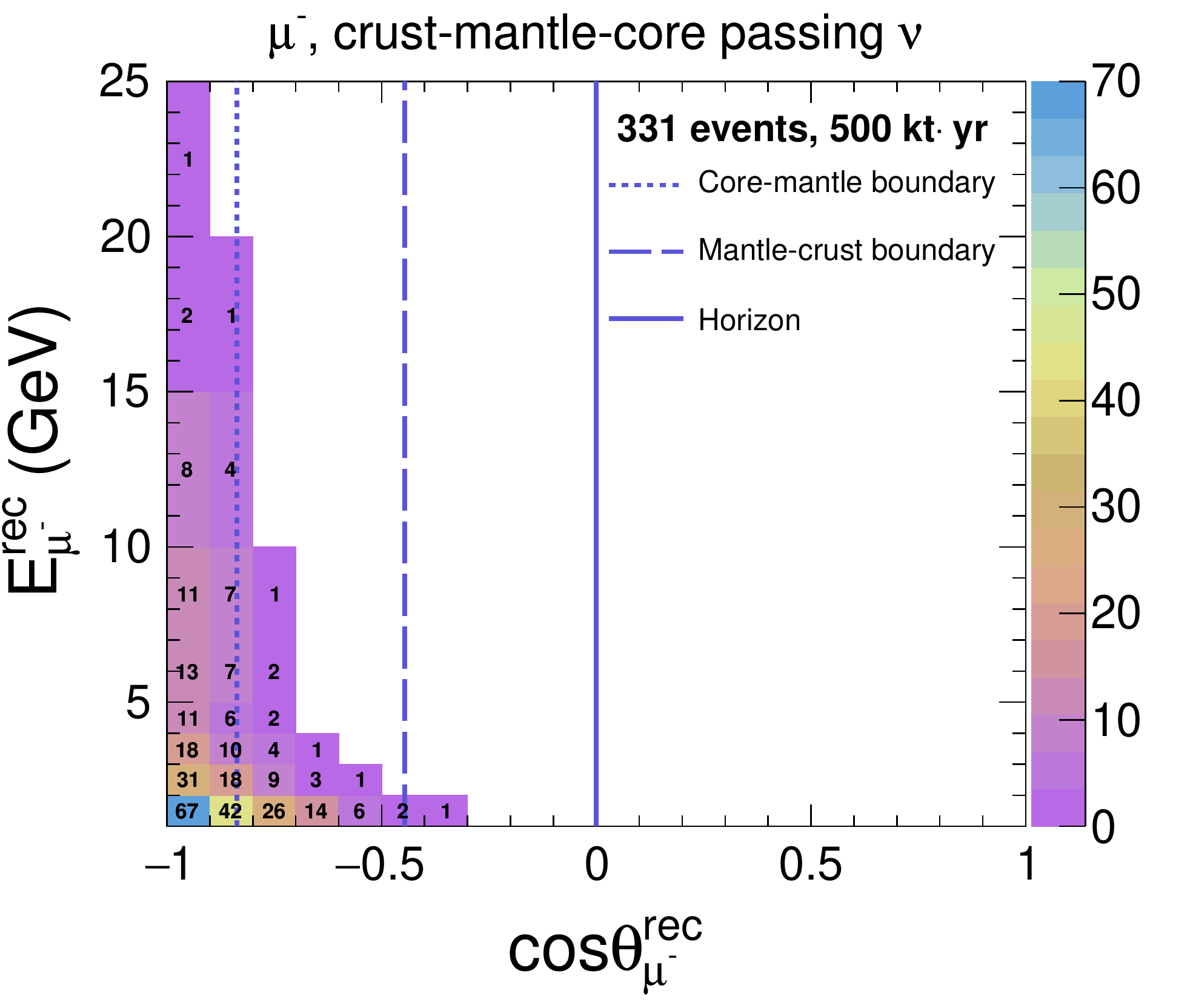} \hspace{0.5cm}
	\includegraphics[width=0.4\linewidth]{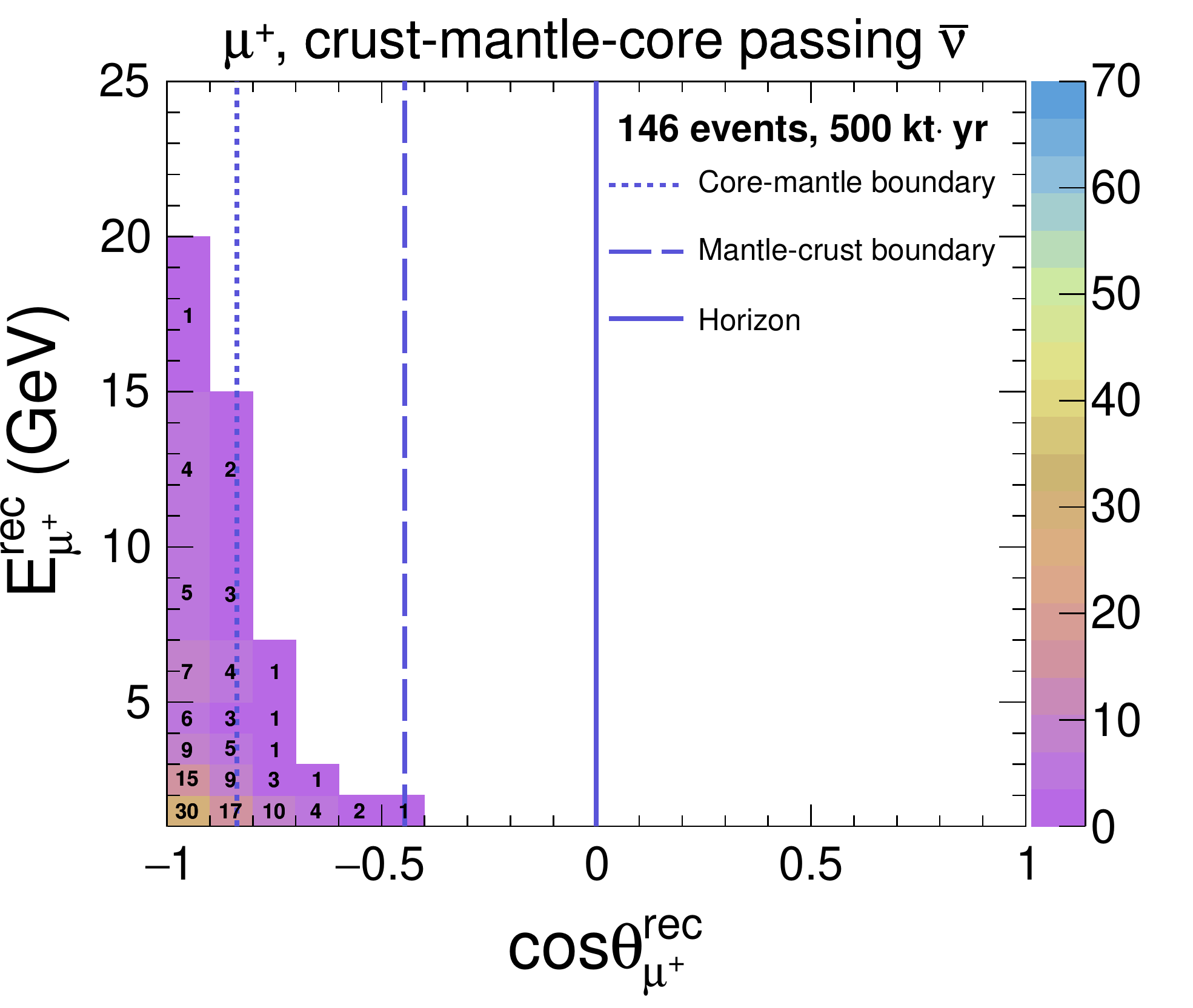}
	\caption{The distributions of reconstructed $\mu^-$ (left panel) and $\mu^+$ (right panel) events with 500 kt$\cdot$yr exposure at ICAL for core-passing neutrinos and antineutrinos, respectively, assuming the PREM profile of Earth. The dotted, dashed, and solid blue vertical lines refer to the core-mantle boundary, mantle-crust boundary, and horizon, respectively. These figures are taken from Ref.~\cite{Kumar:2021faw}.}
	\label{fig:nu-core-mantle-crust}
\end{figure}

Using the good directional resolution, ICAL would be able to detect about 331 $\mu^-$ and 146 $\mu^+$ events for core-passing neutrinos and antineutrinos, respectively, in 10 years. In Fig.~\ref{fig:nu-core-mantle-crust}, we show the distributions of reconstructed $\mu^-$ (left panel) and $\mu^+$ (right panel) events for core-passing neutrinos and antineutrinos, respectively. We can observe that the reconstructed muons preserve the directional information about the regions through which neutrinos have traveled.
 
\section{Results}

To perform numerical analysis, we define a Possonian $\chi^2_{-}$ for $\mu^-$ in terms of reconstructed observables $E_\mu^{\rm rec}$, $\cos\theta_\mu^{\rm rec}$, and ${E'}_{\rm had}^{\rm rec}$ following Refs.~\cite{Devi:2014yaa,Kumar:2021faw}:

\begin{equation}\label{eq:chisq_mu-}
\chi^2_- = \mathop{\text{min}}_{\xi_l} \sum_{i=1}^{N_{{E'}_\text{had}^\text{rec}}} \sum_{j=1}^{N_{E_{\mu}^\text{rec}}} \sum_{k=1}^{N_{\cos\theta_\mu^\text{rec}}} \left[2(N_{ijk}^\text{theory} - N_{ijk}^\text{data}) -2 N_{ijk}^\text{data} \ln\left(\frac{N_{ijk}^\text{theory} }{N_{ijk}^\text{data}}\right)\right] + \sum_{l = 1}^5 \xi_l^2,
\end{equation}
where, 
\begin{equation}
N_{ijk}^\text{theory} = N_{ijk}^0\left(1 + \sum_{l=1}^5 \pi^l_{ijk}\xi_l\right).
\end{equation}
Above, $N_{ijk}^\text{theory}$ and $N_{ijk}^\text{data}$ correspond to the expected and observed number of reconstructed $\mu^-$ events in a given $(E_\mu^{\rm rec},  \cos\theta_\mu^{\rm rec}, {E'}_{\rm had}^{\rm rec})$ bin. We use the binning scheme as mentioned in Table~5 in Ref.~\cite{Kumar:2021faw}. $N_{ijk}^0$ represents the number of events without systematic uncertainties. We incorporate systematic uncertainties using method of pulls as described in Refs.~\cite{Devi:2014yaa,Kumar:2021faw}.

Similarly, we define $\chi^2_+$ for $\mu^+$ and then add $\chi^2_-$ and $\chi^2_+$ to obtain the total $\chi^2_{\rm ICAL}$ for ICAL, 
\begin{equation}
\chi^2_{\rm ICAL} = \chi^2_- (\mu^-) + \chi^2_+ (\mu^+).
\end{equation}
Here, we use the benchmark values of oscillation parameters as mentioned in Table~\ref{tab:osc-param-value} as true parameters to generate the prospective data. In theory, we first minimize $\chi^2_{\rm ICAL}$ with respect to the pull variables ($\xi_l$), and then, we perform marginalization over the oscillation parameters $\sin^2\theta_{23}$ in the range of 0.36 to 0.66, $\Delta m^2_{\rm eff}$ in the range of $(2.1 - 2.6) \times 10^{-3}~{\rm eV}^2$, and both choices of mass orderings (NO and IO). The values of other oscillation parameters in fit are kept fixed at their benchmark values as mentioned in Table~\ref{tab:tomography_results}.

To perform statistical analysis, we first simulate the prospective data assuming the PREM profile as the true profile of the Earth. The statistical significance of ICAL to rule out the mantle-crust profile with respect to the PREM profile is defined as follows
\begin{equation}\label{eq:chisq_diff}
\Delta \chi^2_\text{ICAL-profile} = \chi^2_\text{ICAL}~ (\text{mantle-crust}) - \chi^2_\text{ICAL}~ (\text{PREM}),
\end{equation} 
where $\chi^2_\text{ICAL}~ (\text{mantle-crust})$ and $\chi^2_\text{ICAL}~ (\text{PREM})$ are estimated by fitting the prospective data assuming mantle-crust and PREM profiles, respectively. Note that $\chi^2_\text{ICAL}~ (\text{PREM}) \approx 0$ because the statistical fluctuations are suppressed to obtain the median sensitivity.

\begin{table}
	\centering
	\begin{tabular}{|c|c|c|c|c|c|}
		\hline \hline
		\multirow{3}{*}{MC Data} & \multirow{3}{*}{Theory} & \multicolumn{4}{c|}{$\Delta \chi^2_\text{ICAL-profile}$}\\ \cline{3-6}
		& & \multicolumn{2}{c|}{NO(true)} & \multicolumn{2}{c|}{IO(true)}\\ \cline{3-6}
		& & CID & No CID & CID & No CID\\
		\hline
		PREM & Vacuum & 5.52 & 3.52 & 4.09 & 1.67 \\
		PREM & Mantle-Crust & 7.45 & 3.76 & 4.83 & 1.59 \\
		PREM & Core-Mantle & 0.27 & 0.18 & 0.21 & 0.07 \\
		PREM & Uniform  & 6.10 & 3.08 & 3.92 & 1.18 \\
		\hline \hline
	\end{tabular}
	\caption{The median $\Delta\chi^2$ to rule out the alternative profiles of Earth with 500 kt$\cdot$yr exposure of ICAL. We perform marginalization over oscillation parameters $\sin^2 \theta_{23}$, $\Delta m^2_{\rm eff}$, and both the mass orderings in theory while keeping other oscillation parameters fixed at their benchmark values as mentioned in Table~\ref{tab:osc-param-value}. These sensitivities are taken from Ref.~\cite{Kumar:2021faw}.
	}
	\label{tab:tomography_results}
\end{table}

In Table~\ref{tab:tomography_results}, we present ICAL sensitivities to rule out the alternative profiles of Earth in fit with respect to the PREM profile in MC data. We can observe that $\Delta \chi^2_\text{ICAL-profile}$ to rule out the vacuum in theory with respect to the PREM profile in MC data is 5.52 for NO (true) and 4.09 for IO (true) with CID, which shows that ICAL can establish the presence of Earth's matter with good sensitivity. In the absence of CID, these results deteriorate to 3.52 for NO (true) and 1.67 for IO (true), which indicate that the CID capability of ICAL is crucial to observe the Earth's matter effect.

The second row in Table~\ref{tab:tomography_results} shows that ICAL can rule out the coreless profile of mantle-crust in fit with respect to the PREM profile in MC data with $\Delta \chi^2_\text{ICAL-profile}$ of 7.45 for NO (true) and 4.83 for IO (true) with CID. This is the sensitivity with which ICAL can validate the presence of Earth's core. In the absence of CID, these sensitivities worsen to 3.76 for NO (true) and 1.59 for IO (true), which shows that CID plays a vital role in establishing Earth's core using ICAL. We also observe that ICAL would be able to validate Earth's core with a higher confidence level if $\theta_{23}>45^\circ$.

\section{Conclusions}

The atmospheric neutrinos passing through Earth interact with the ambient electrons, which modifies the neutrino and antineutrino oscillation probabilities in different ways. These density-dependent matter effects can be utilized to reveal the distribution of matter inside Earth. Using the good energy and directional resolutions, the 50 kt ICAL would be able to observe 331 $\mu^-$ and 146 $\mu^+$ events for core-passing neutrinos and antineutrinos, respectively, in 10 years. We demonstrate that using 500 kt$\cdot$yr exposure, ICAL would be able to validate the presence of Earth's core by ruling out the coreless profile of mantle-crust in theory with respect to the PREM profile in data with a $\Delta \chi^2$ of 7.45 for NO and 4.83 for IO. In the absence of CID, these sensitivities deteriorate to 3.76 for NO and 1.59 for IO.

\textbf{Acknowledgements:} We acknowledge financial support from the DAE,
DST, DST-SERB, Govt. of India, and INSA.

\end{document}